\documentclass[manuscript]{aastex}

\usepackage{natbib}

\newcommand{\teff}{$T_\mathrm{eff}$}
\newcommand{\logg}{$\log g$}
\newcommand{\feh}{[Fe/H]}

\newcommand{\kms}{km\,s$^{-1}$}
\newcommand{\mic}{$\mu \mathrm m$}

\shorttitle{Chemical evolution of the inner Bulge}
\shortauthors{Ryde et al.}

\begin{document}


\title{Chemical evolution of the inner 2 degrees of the Milky Way bulge:\\
$[\alpha/\mathrm{Fe}]$ trends and metallicity gradients
}

\author{N. Ryde\altaffilmark{1}}
\affil{Department of Astronomy and Theoretical Physics, Lund Observatory, Lund University, Box 43, SE-221 00, Lund, Sweden}
\email{ryde@astro.lu.se}

\author{M. Schultheis\altaffilmark{1}}
\affil{Laboratoire Lagrange, Universit\'e C\^ote d'Azur, Observatoire de la C\^ote d'Azur, CNRS, Blvd de l'Observatoire, F-06304 Nice, France}

\author{V. Grieco and F. Matteucci\altaffilmark{2,3}}
\affil{Dipartimento di Fisica, Sezione di Astronomia, Universit\`a di Trieste, via G.B. Tiepolo 11, I-34131, Trieste, Italy}

\author{R. M. Rich}
\affil{Department of Physics and Astronomy, UCLA, 430 Portola Plaza, Box 951547, Los Angeles, CA 90095-1547, USA}

\and

\author{S. Uttenthaler}
\affil{University of Vienna, Department of Astrophysics, T\"urkenschanzstra\ss e 17, A-1180 Vienna, Austria}


\altaffiltext{1}{Visiting Astronomer, VLT, the European Southern Observatory, Chile. 
}
\altaffiltext{2}{I.N.A.F. Osservatorio
  Astronomico di Trieste, via G.B. Tiepolo 11, I-34131, Trieste,
  Italy}
\altaffiltext{3}{I.N.F.N. Sezione di Trieste, via Valerio 2, I-34134 Trieste, Italy }


\begin{abstract}
The structure, formation, and evolution of the Milky Way bulge is a matter of debate. Important diagnostics for discriminating between models of bulge formation and evolution include  $\alpha$-abundance trends with metallicity, and spatial abundance and metallicity gradients. Due to the severe optical extinction in the inner Bulge region, only a few detailed investigations have been performed of this region.
   Here we aim at investigating the inner 2 degrees of the Bulge (projected galactocentric distance of approximately 300 pc), rarely investigated before, by observing the [$\alpha$/Fe] element trends versus metallicity, and by trying to derive the metallicity gradient in the $b<2^\circ$ region. 
   [$\alpha$/Fe] and metallicities have been determined by spectral synthesis of 2\,\mic\ spectra  of 28 M-giants in the Bulge, lying along the Southern minor axis at $(l,b)=(0,0)$, $(0,-1^\circ)$, and $(0,-2^\circ)$. These were observed with the CRIRES spectrometer at the {\it Very Large Telescope, VLT} at high spectral resolution. Low-resolution K-band spectra, observed with the ISAAC spectrometer at the {\it VLT}, are used to determine the effective temperature of the stars. 
   We present the first connection between the Galactic Center and the Bulge using
similar stars, high spectral resolution, and analysis techniques. The [$\alpha$/Fe] trends in all our three fields show a large similarity among each other and with trends further out in the Bulge. All point to a rapid star-formation episode in the Bulge. We find that there is a lack of an [$\alpha$/Fe] gradient in the Bulge all the way into the centre, suggesting a homogeneous Bulge when it comes to the enrichment process and  star-formation history. We find a large range of metallicities from $-1.2<\mathrm{[Fe/H]}<+0.3$, with a lower dispersion in the Galactic center: $-0.2<\mathrm{[Fe/H]}<+0.3$. The derived metallicities of the stars in the three fields get, in the mean, progressively higher the closer to the Galactic plane they lie. 
   We could interpret this as a continuation of the metallicity gradient established further out in the Bulge, but due to the low number of stars and possible selection effects, more data of the same sort as presented here is necessary to conclude on the inner metallicity gradient from our data alone.  
   Our results firmly argues for the center being in the context of the Bulge rather than very distinct. 
\end{abstract}





\section{Introduction}

The current knowledge and main ideas of the structure, formation, evolution of the Milky Way bulge are under debate and in rapid development, mainly due to the recent large amount of new observations of  kinematics, photometry, and abundances of Bulge stars, as well as sophisticated modelling and interpretation. The Milky Way bulge or bulge/bar can nowadays be seen upon as the inner structure of the bar \citep{gerhard:14} and  is certainly a complex environment. It might be a composite system with a mixture of different populations, similarly to many bulges at low redshift \citep{sanchez:15}. For recent reviews of the Milky Way bulge, see \citet{rich:13},  \citet{gonzalez:15}, and \citet{shen:15}.

Originally, it was thought that the Milky Way bulge was a classical bulge formed by an initial collapse, with an early, rapid, and intense star formation, in an inside-out formation scenario \citep[e.g.][]{eggen:62}, or through accretion of substructures and/or large violent merger events in the $\Lambda$CDM scenario \citep[e.g.][]{immeli:04,shen:15}. However, the observational evidence for a classical bulge has weakened and the classical contribution seems to be small \citep{shen:10,dimatteo:15}, see however the discussions in \citet{hill:11} and \citet{saha:15}. 


At present, a range of observations points toward a complex situation which may be consistent with multiple components/populations or gradients within populations \citep[e.g.][]{hill:11,ness:13,dekany:13,rojas:15}.  On the one hand, kinematic surveys, e.g. \cite{kunder:12}
and \citet{ness:13} are strongly consistent with the bar/X-structure being the predominant population by mass \citep[see also][]{wegg:13}. However, there are strong correlations observed between abundances and kinematics, which pose challenges to any unified picture \citep[e.g.][]{babu:10,johnson:11,johnson:2014,ness:13}.  According to the chemical evolution models of  \citet{grieco:12}, the modelled spheroidal component formed rapidly on a short timescale via an intense burst of star formation in a classical gravitational gas collapse, and the component created by bar evolution formed later on  a longer  time scale. 
Note, however, that much of the metal-rich component has elevated alphas \citep[][see also this work]{rich:12,johnson:2014}.
The nature of these components are not certain \citep[see e.g.][]{johnson:2014} and associating them with specific formation scenarios should be done with caution \citep{gonzalez:15}.  

Important diagnostics for discriminating between models of bulge formation and evolution are   {\it (i)} $\alpha$-abundance trends with metallicity, and {\it (ii)} spatial abundance-gradients in the Bulge:

{\it (i)}  Abundance trends can set constraints on the formation time-scales of the different components \citep{matteucci:12}.  
 The dynamics and abundance trends of Bulge stars from the metal-poor component \citep[even though there are differences, see e.g.][]{johnson:2014} point to similarities especially with the (local) thick disk \citep[e.g.][]{melendez:2008,hill:11}. \citet{grieco:12}  show that from a chemical point of view, an inside-out formation, modelled with a initial collapse, be it a due to an early star burst in a disk,  a classical bulge, or a thick disk, fits the chemical data. Large samples with chemistry and dynamics are  needed in order to constrain the formation in detail. 
 

{\it (ii)} Abundance gradients are important outcomes of formation models. There is observational evidence that there is a metallicity gradient from $b=-12^\circ$ to $-4^\circ$  latitude \citep[e.g.][]{zoccali:2003,johnson:11,johnson:13,gonzalez:13,rojas:15}, but the origin of this is debated. It could either reflect varying relative proportions of the different components \citep{babu:10,gonzalez:11,rojas:15} or it could reflect an intrinsic metallicity gradient produced by the bar due to an original disk gradient \citep{gerhard:13}. 
Within $2^\circ$, only a few investigations  of the metallicity distribution based on stellar spectra at medium- and high resolutions exist. The Galactic centre (GC) and the region defined by the inner few hundred parsecs have escaped detailed study of  stellar abundances and abundance gradients from giants, due to the extreme optical obscuration, resulting in only a handful studies in the literature. \citet{rich:07,rich:12}  provide evidence for a break in the gradient, being absent with a narrow metallicity distribution at [Fe/H]$\sim -0.1$ between $b=-3^\circ$ and $-1^\circ$ to the South of the Galactic Centre. These studies are based on $R\sim20,000$ spectroscopy of 50 giants observed in the H band. Recently, \citet{massari:14} found a double-peaked distribution with peaks at slightly sub-solar and super-solar metallicities
at a Northern latitude of $b=+1.7^\circ$, based on medium-resolution spectra of 112 stars in the field. \citet{ramirez:00} and \citet{cunha2007} obtained K-band spectra of the supergiant population in the 
Galactic center, finding [Fe/H]$=+0.1$, with a narrow dispersion. Recently, \citet{ryde_schultheis:15} also find a metallicity peak in the Galactic centre at [Fe/H]$\sim+0.1$, but with a slightly larger spread, also from K-band spectra. All these Galactic Centre studies are based on approximately 10 stars each, so caution should be exercised when drawing conclusions. Recently, \citet{do:15} determined metallicities of 83 giants within the inner 1 pc of the Galactic center, based on low-resolution IFU spectra ($R=5000$), and found an surprisingly large dispersion ($-1.3<$[Fe/H]$<0.9$). High-resolution spectra are needed to validate these results.


In this paper, we  will discuss these issues for the inner bulge, within a projected distance of 300 pc ($-2^\circ<b<0^\circ$) from the center of the Galaxy. 
To overcome the extreme optical extinction in these regions, we observe all our stars at $2.0\,$\mic. The extinction in the K band is a factor of 10 lower that in the V band \citep{cardelli}. The large crowding of stars can be dealt with thanks to Adaptive Optics. To ensure a detailed abundance analysis, we record our spectra with high spectral resolution. We will be able to link the Galactic center and the bulge by analysing similar stars and using the same analysis techniques including the same spectral region and high spectral resolution, which to our knowledge has not really been done before. The luminosity and temperature ranges of our stars in the Galactic center and the other two fields is similar as well. As presented in \citet{ryde_schultheis:14}, we also argue that this population is at least 1 Gyr old, with no substantial membership from the youngest 10 Myr stars.



\section{Observations} \label{observations}

\begin{figure*}[!h]
  \centering
	\includegraphics[width=0.33\textwidth]{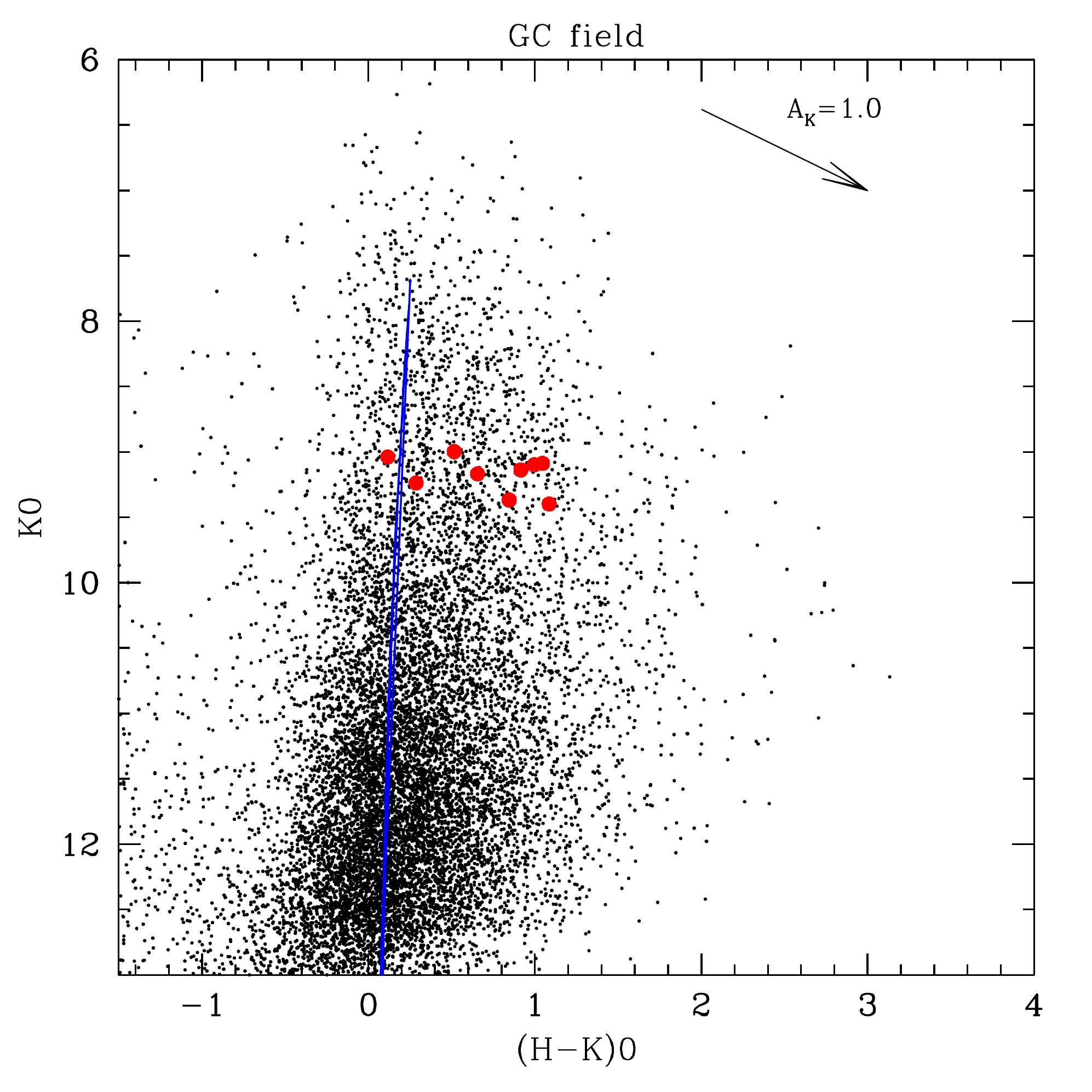} \includegraphics[width=0.33\textwidth]{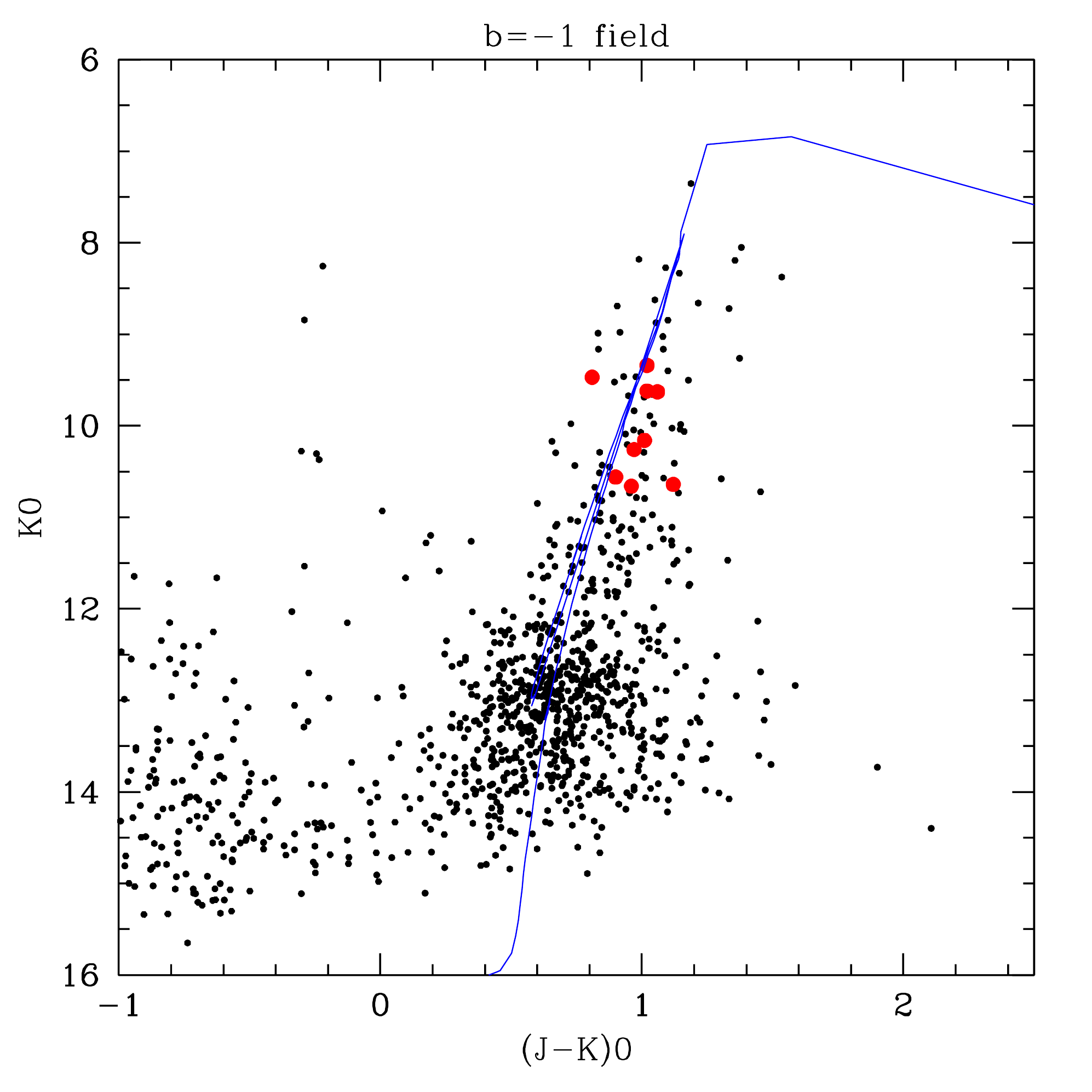}\includegraphics[width=0.33\textwidth]{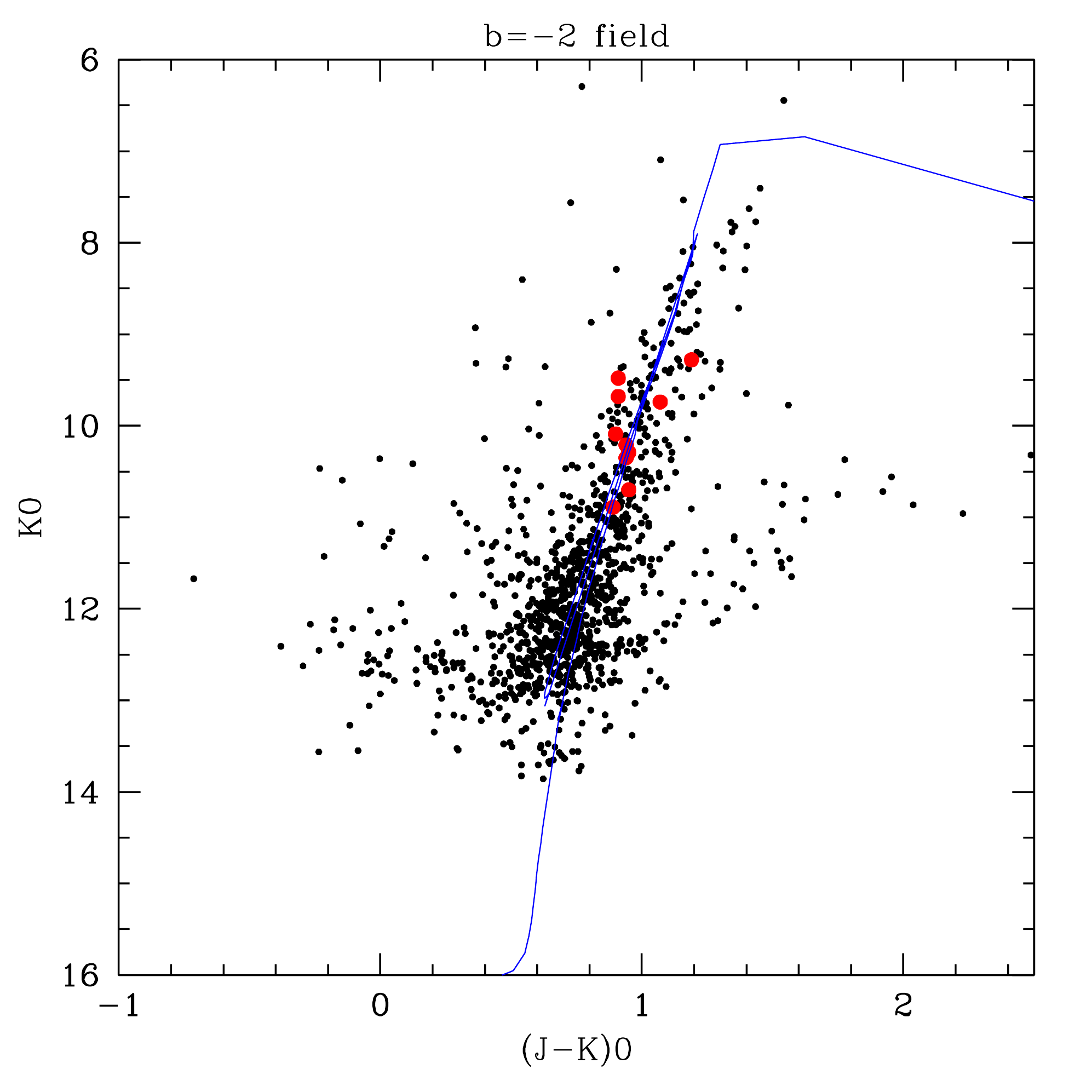}
	\caption{Color-magnitude diagrams for dereddened magnitudes of stars  
	in the Galactic Center field, the $b=-1^\circ$, and $b=-2^\circ$ fields, based on data from \citet{nishiyama2009}  and  2MASS \citep{2mass} and using the extinction map
of \citet{schultheis2009}.  Red dots mark our selected M giants. The blue lines show the Padova $\rm  10\,Gyr$ isochrones. The arrow on the top right of the Galactic Center panel, shows the extinction vector of $\rm A_{K} = 1\,mag$ \citep{nishiyama2009}.
	} 
	\label{cmdall}
\end{figure*}

\begin{deluxetable}{l c c c c c c c c c c r}
\tabletypesize{\scriptsize}
\tablecaption{Basic data for the observed stars and their stellar parameters.\label{starsall_tab}}
\tablewidth{0pt}
\tablehead{
\colhead{Star} & \colhead{RA (J2000)} & \colhead{Dec (J2000)} & \colhead{$H$} & \colhead{$K_s$} &
\colhead{$H_0$} & \colhead{$K_0$} & \colhead{exp. time} &
\colhead{\teff}  & \colhead{\logg} & \colhead{\feh}  & \colhead{$\xi_\mathrm{macro}$}\\
\colhead{} & \colhead{(h:m:s)} & \colhead{(d:am:as)} & \colhead{} & \colhead{} &
\colhead{} & \colhead{} & \colhead{[s]} &
\colhead{[K]} & \colhead{(cgs)} & \colhead{} & \colhead{km\,s$^{-1}$  } 
}
%
\startdata
\multicolumn{4}{l}{Galactic centre stars at $(l,b) = (0^{\circ},0^{\circ})$}\\
\tableline
GC1   & 17:45:35.43 &  -28:57:19.28 & 14.74 & 11.90  &  10.48 & 9.40 & 3000 & 3668 & 1.37 & $+$0.15   & 6.6   \\
GC20  & 17:45:34.95 &  -28:55:20.17 & 14.47 & 11.87  &  10.21 & 9.37 & 3600 & 3683 & 0.76 & $+$0.14   & 8.1    \\
GC22  & 17:45:42.41 &  -28:55:52.99 & 13.41 & 11.54  &  9.15 & 9.04 & 3600 & 3618 & 0.70 & $+$0.04   & 6.0    \\
GC25  & 17:45:36.34 &  -28:54:50.41 & 14.35 & 11.60  &  10.09 & 9.10 & 2400 & 3340 & 0.77 & $-$0.20   & 7.6     \\
GC27  & 17:45:36.72 &  -28:54:52.37 & 14.31 & 11.64  &  10.05 & 9.14 & 3600 & 3404 & 1.16 & $+$0.23   & 8.1       \\
GC28  & 17:45:38.13 &  -28:54:58.32 & 14.08 & 11.67  &  9.82 & 9.17 & 3000 & 3773 & 1.24 & $-$0.04   & 5.7        \\
GC29  & 17:45:43.12 &  -28:55:37.10 & 14.39 & 11.59  &  10.13 & 9.09 & 3600 & 3420 & 0.74 & $+$0.12   & 8.0        \\
GC37  & 17:45:35.94 &  -28:58:01.43 & 13.77 & 11.50  &  9.51 & 9.00 & 3600 & 3754 & 0.93 & $-$0.08   & 7.7      \\
GC44  & 17:45:35.95 &  -28:57:41.52 & 13.78 & 11.74  &  9.52 & 9.24 & 3600 & 3465 & 0.83 & $+$0.18   & 6.9      \\
\tableline
\multicolumn{4}{l}{Bulge stars at $(l,b) = (0^{\circ},-1^{\circ})$}\\
\tableline
bm1-06 & 17:49:33.42 & -29:27:28.75 & 11.84 & 11.04 & 10.58 & 10.26 & 1800 &  3814 & 	1.18 & 	 0.29  & 5.1	\\
bm1-07 & 17:49:34.58 & -29:27:14.82 & 12.22 & 11.44 & 10.95 & 10.66 & 2400 &  3873 & 	1.34 & 	 0.08  & 4.2	\\
bm1-08 & 17:49:34.34 & -29:26:57.98 & 11.22 & 10.40 &  9.95 &  9.62 & 1200 &  3650 & 	0.89 & 	 0.17  & 4.1	\\
bm1-10 & 17:49:34.45 & -29:26:48.68 & 10.92 & 10.12 &  9.65 &  9.34 & 1920 &	3787 & 		0.80 & 	-0.23  & 5.0	\\
bm1-11 & 17:49:32.57 & -29:26:30.75 & 11.26 & 10.41 & 10.00 &  9.63 & 1200 &  3812 & 	0.92 & 	 0.12  & 4.6	\\
bm1-13 & 17:49:37.12 & -29:26:40.24 & 10.91 & 10.25 &  9.65 &  9.47 & 1080 &	3721 & 		0.84 & 	-0.94  & 4.9	\\
bm1-17 & 17:49:37.08 & -29:26:21.67 & 11.70 & 10.94 & 10.43 & 10.16 & 1800 &	3775 & 		1.13 & 	-0.83  & 5.0	\\
bm1-18 & 17:49:37.83 & -29:26:19.19 & 12.23 & 11.42 & 10.96 & 10.64 & 1800 &	3780 & 		1.32 & 	 0.22 & 2.7 \\	
bm1-19 & 17:49:36.93 & -29:26:10.51 & 12.11 & 11.34 & 10.84 & 10.56 & 3600 &	3958 & 		1.32 & 	 0.18  & 4.0 \\
\tableline
\multicolumn{4}{l}{Bulge stars at $(l,b) = (0^{\circ},-2^{\circ})$}\\
\tableline
bm2-01 & 17:53:29.06 & -29:57:46.22 & 11.44 & 11.11 & 11.08 & 10.89 & 2400 & 	3946 & 	1.00 & 	 0.14	 & 4.2	\\
bm2-02 & 17:53:24.59 & -29:59:09.48 & 10.78 & 10.44 & 10.41 & 10.21 & 1920 & 	4013 & 	1.19 & 	-0.48	 & 5.6	\\
bm2-03 & 17:53:27.61 & -29:58:36.39 & 11.30 & 10.93 & 10.94 & 10.70 & 2400 & 	3668 & 	1.33 & 	 0.26	 & 3.5 	\\
bm2-05 & 17:53:33.20 & -29:57:25.88 & 10.07 &  9.51 &  9.70 &  9.28 & 2280 &	 3450 & 	0.72 & 	 0.01	 & 4.2	\\
bm2-06 & 17:53:30.68 & -29:58:15.75 & 10.01 &  9.71 &  9.65 &  9.48 & 1200  & 	4208 & 	0.92 & 	-1.17   & 5.6 \\
bm2-11 & 17:53:31.50 & -29:58:28.51 & 10.20 &  9.91 &  9.83 &  9.68 & 1680  & 	4005 & 	0.98 & 	-0.91	 & 5.8	\\
bm2-12 & 17:53:31.74 & -29:58:22.94 & 10.67 & 10.32 & 10.30 & 10.09 & 480  & 	4003 &  1.14 & 	-0.11	 & 6.2	\\
bm2-13 & 17:53:31.14 & -29:57:32.76 & 10.86 & 10.52 & 10.50 & 10.29 & 2400 & 	3727 & 	1.18 & 	-0.16	 & 5.2	\\
bm2-15 & 17:53:30.23 & -29:56:42.74 & 10.43 &  9.96 & 10.07 &  9.74 & 1200 & 3665 & 	1.01 & 	 0.22	 & 4.5	\\
bm2-16 & 17:53:29.54 & -29:57:22.71 & 10.93 & 10.57 & 10.56 & 10.35 & 1200 & 3886 & 	1.22 & 	 0.10	 & 3.2	\\
\tableline
\multicolumn{4}{l}{Disk stars}\\
\tableline
$\alpha$ Boo & 14:15:39.67 & $+$19:10:56.7 & $-$2.81 & $-$2.91 & & & $\mathrm{atlas}$ & 4286 & 1.66 & $-$0.52  & 6.3    \\ 
BD-012971     & 14:38:48.04 & $-$02:17:11.5 & 4.59 & 4.30 & & &  16 & 3573 & 0.50 &  $-$0.78   & 9.3 \\ 
142173$^{a}$  & 00:32:12.56 & $-$38:34:02.3 & 9.40 & 9.29 & & & 240 & 4330 & 1.50 &  $-$0.77   & 9.3 \\ 
171877$^{a}$  & 00:39:20.23 & $-$31:31:35.5 & 8.25 & 8.12 & & & 240 & 3975 & 1.10 &  $-$0.93   & 7.0 \\ 
225245$^{a}$  & 00:54:46.38 & $-$27:35:30.4 & 8.90 & 8.79 & & & 240 & 4031 & 0.65 &  $-$1.16   & 6.6 \\ 
313132$^{a}$  & 01:20:20.66 & $-$34:09:54.1 & 7.22 & 7.04 & & & 120 & 4530 & 2.00 &  $-$0.20   & 8.7 \\ 
343555$^{a}$  & 01:29:42.01 & $-$30:15:46.4 & 8.59 & 8.46 & & & 240 & 4530 & 2.25 &  $-$0.74   & 6.8 \\ 
HD 787        & 00:12:09.99 & $-$17:56:17.8 & 2.03 & 1.86 & & & 4 & 4020 & 0.85 &  $-$0.16     & 8.2 \\ 

\enddata
\tablenotetext{a}{Identification number from the Southern Proper-Motion Program \citep[SPM III][]{girard:04}, as given in \citet{monaco:11}.}
\end{deluxetable}

During the three nights of June, 27-29, 2012, we observed 28 Bulge giants within 2 degrees (approximately 300 pc in projected distance) from the Galactic center, along the Southern minor axis, using the {\it Very Large Telescope, VLT,} spectrometers CRIRES \citep{crires} and ISAAC \citep{isaac}. We observed 9 stars at  the Galactic center (GC), 
 9 stars at $(l,b) = (0,-1^{\circ})$, and 10 stars at  $(l,b) = (0,-2^{\circ})$, apart from 7 disk stars which are used for comparison. All the stars are M giants (\teff$=3300-4200$~K and  0.7 $< \log g <$ 2.25), for which we will determine the metallicities, [Fe/H], and the abundances of the $\alpha$ elements Mg and Si. Typical magnitudes of our Bulge stars are $K_s = 9.5-12.0$. The stars are presented in Table \ref{starsall_tab}, as well as the six local thick-disk giants, the thin-disk giant BD-012971, and the reference star $\alpha$ Boo, which we also will analyse in the same way as for the Bulge stars. The $H$ and $K_s$ magnitudes of the GC stars are from Nishiyama (private comm.) and the stars in the  $b=-1^\circ$ and $b=-2^\circ$ fields, as well as the disk stars, are from 2MASS \citep{2mass}.

Our Bulge stars were selected in the following way:  For the GC and the $b=-1^\circ$ fields, we used the \citet{nishiyama2009}  data set while for the $b=-2^\circ$ field the 2MASS
data set \citep{2mass}. The stars are chosen in colour-magnitude diagrams that were dereddened using the high-resolution 3D interstellar extinction map of \citet{schultheis:14} and assuming the extinction law of \citet{nishiyama2009}. The three panels in Figure \ref{cmdall} show the corresponding dereddened colour magnitude diagram for the GC, $b=-1^\circ$ and $b=-2^\circ$ fields, respectively. Note that due to the extreme interstellar extinction in the Galactic center, the errors in the dereddening procedure remain large which is seen in the larger dispersion of the CMD diagram of the Galactic center compared to the $b=-1^\circ$ and $b=-2^\circ$ fields. In addition, due to the high interstellar absorption in the Galactic center, we use  the $ (H-K)_{0}$ vs. $K_{0}$ diagram, since even the $J$ magnitudes are not useful due to extinction.

 Our stars were chosen such, that they cover the full colour range in
$(J-K)_{0}$ or $(H-K)_{0}$ across the RGB/AGB to avoid any biases against metal-rich or metal-poor stars.  However, in the Galactic center
region, small-scale differential extinction not resolved by extinction maps, could lead to possible biases against metal-poor stars. These biases are difficult to quantify, but should be kept in mind. In Figure \ref{cmdall} our selected stars are shown with red dots superimposed by a few Padova isochrones \citep{bressan:12}. Evidently, the $H-K$ colour is insensitive to the age of the population, prohibiting an age constrain and therefore a constraint on the mass of our stars. The location of our stars in the CMD shows that we are dealing with M giants and not supergiants.

\begin{figure}
	\includegraphics[width=\hsize]{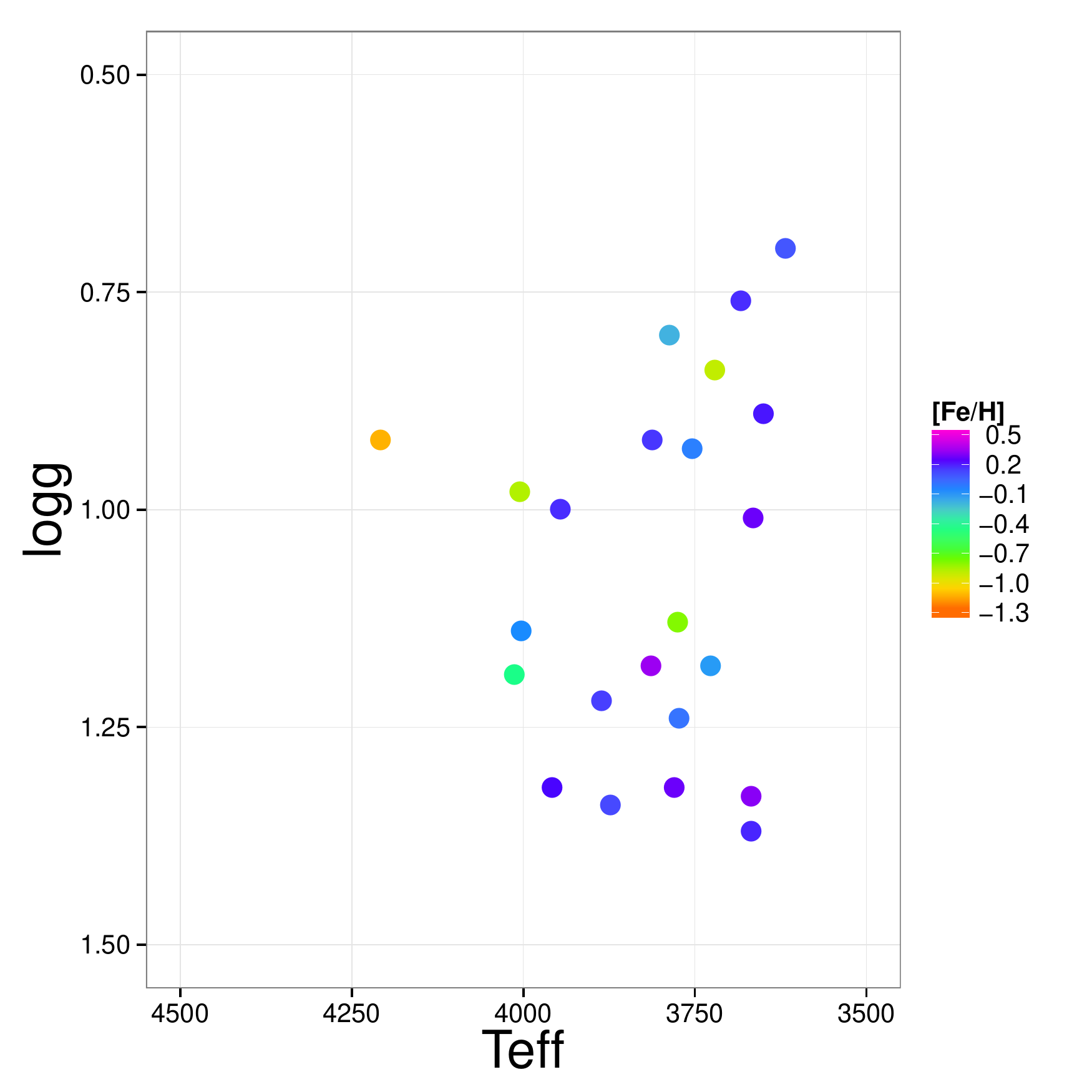}
	\caption{Color-Magnitude diagram for the stars in all our three fields. The stars are color coded according to their metallicities. The \teff, \logg, and metallicity are determined from our spectra, see  Sect. \ref{param}.} 
	\label{cmd}
\end{figure}

To ensure that
our sources are located  in the Bulge, we placed our stars on the 3D extinction plots of \citet{schultheis:14} and found that our stars lie between $7-9$\,kpc which rules out any possible foreground contamination.  Furthermore, our derived radial velocity dispersion is about  $\rm 150\pm40\,km\,s^{-1}$,
which is consistent with a  typical velocity dispersion of the 
Bulge kinematics \citep[e.g.][]{rich:07}.
Figure \ref{cmd} shows the Hertzsprung-Russell diagram  of our
selected sources as a function of metallicity. It is based on our derived effective temperatures and surface gravities, $\log g$, see Section \ref{param}. It represents nicely the RGB branch covering a wide  range of metallicities  ($-1.3 < $[Fe/H]$ < +0.5$).  

To overcome the extreme extinction we thus observed the giants with the near-IR spectrometers CRIRES \citep{crires} mounted on UT1 (Antu), and ISAAC \citep{isaac} on UT3 (Melipal) of the {\it VLT}. Every star was observed with both spectrometers. 
The ISAAC spectrometer, with a spectral resolution of  $R\sim1000$ and a wavelength range between $2.00-2.53$\,\mic, was used to determine the effective temperature of our stars.  

Typical integration times for the high-resolution CRIRES spectra ($R =  50,000$, resulting from the $0.''4$ wide slit) are $1/2 - 1$ hour per star (see Table \ref{starsall_tab}), giving a signal-to-noise ratios per pixel of typically 50-100. CRIRES uses, following standard procedures \citep{crires:manual}, nodding on the slit and jittering to reduce the sky background and the Adaptive Optics (AO) MACAO system to enhance the amount of light captured by the slit. For the CRIRES observations, we used a standard setting ($\lambda^\mathrm{ref}_\mathrm{vac}=2105.5$, order=27) with an unvignetted spectral range covering $20818-21444$\,\AA, with three gaps (20\,\AA) between the four detector arrays.

\begin{figure*}
  \centering
	\includegraphics[angle=90,width=\textwidth]{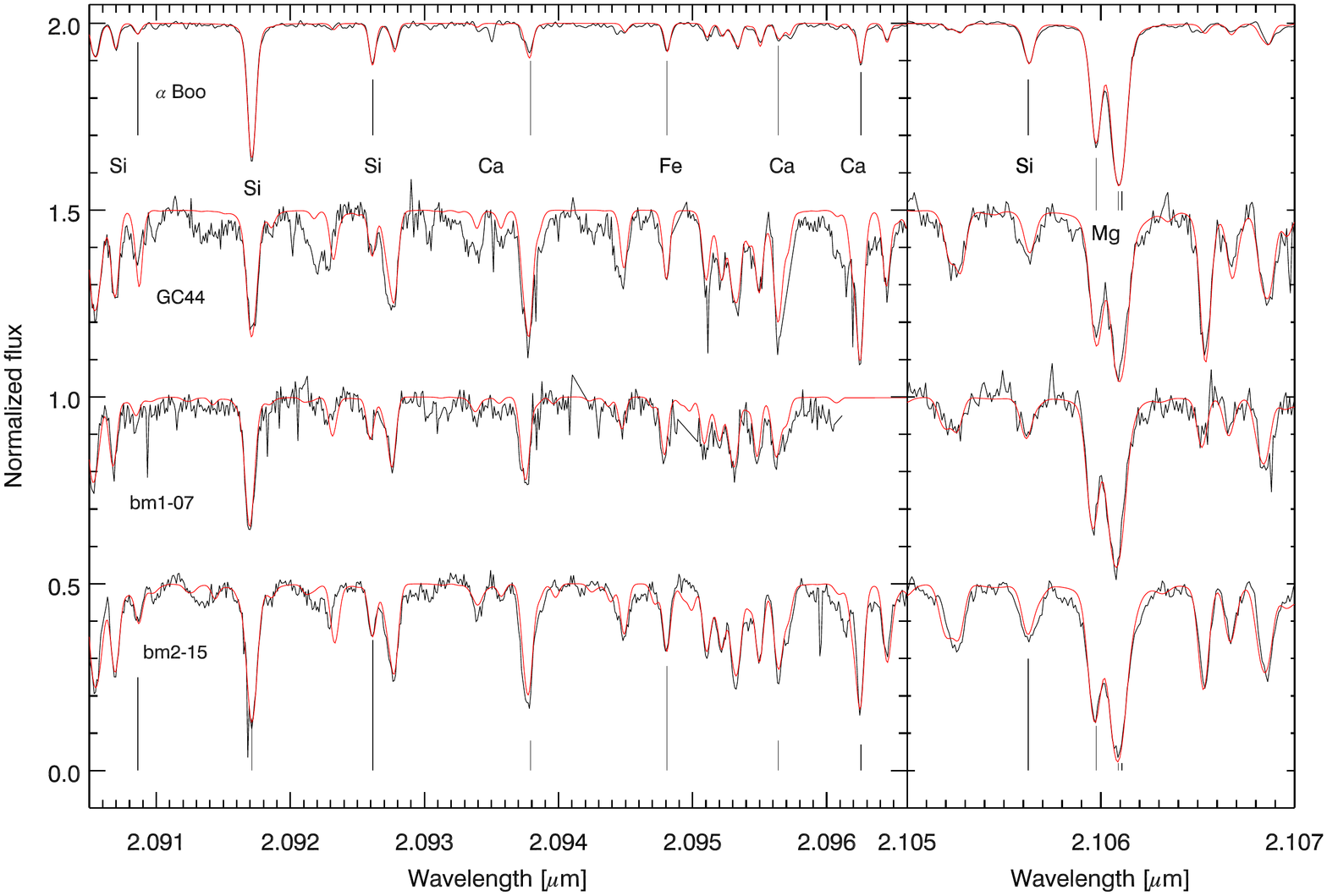}
	\caption{Spectra of wavelength regions covering a few of the lines used for the abundance determination. Other features not marked are mostly due to CN. The top spectrum is from the Arcturus atlas \citep{arcturusatlas_II}, the next one is of the Galactic center star GC44, and the bottom two spectra are from the $(l,b) = (0,-1^{\circ})$ (bm1-07) and $(0,-2^{\circ})$  (bm2-15) fields, respectively. The stellar parameters (\teff,\logg, and [Fe/H]) of the stars are, from top to bottom, (4286,1,66,$-0.52$), (3465,0.83, $+0.18$), (3873,1.34,$+0.08$), and (3665,1.01,$+0.22$).} 
	\label{spectra}
\end{figure*}

The reductions of the ISAAC and CRIRES observations were done by following standard recipes \citep{isaac:cook,crires:cook}. For the reduction of the CRIRES spectra, we used Gasgano \citep{gasgano} and, subsequently, IRAF \citep{IRAF} to normalise the continuum, eliminate obvious cosmic hits and correct for telluric lines (with telluric standard stars). Two examples of CRIRES spectra are shown in Figure \ref{spectra}.





\section{Analysis}

We have derived the metallicity and the abundances of the $\alpha$ elements Mg and Si for all our stars. Here, we describe how we determine the stellar abundances and the important stellar parameters. i.e. the effective temperature (\teff), the surface gravity (\logg), the metallicity ([Fe/H]), and the microturbulence ($\chi_\mathrm{micro}$), which are needed in order to derive the abundances.

\subsection{Stellar parameters\label{param}}

We have determined the effective temperatures of our stars in the same manner as in \citet{blum2003}, using the fact that the CO-band-head at $2.3\,$\mic, in our type of stars, observed at low-resolution, is very temperature sensitive, more so than to any other parameter. \citet{ramirez1997}, \citet{ivanov2004}, and  \citet{blum2003}  studied  the behaviour of the $^{12}$CO band-head with low-resolution, K-band spectra and found  a remarkably  tight relation between the equivalent width of the CO band-head and the effective temperature for M giants. We have thus used this property and obtained relevant low-resolution spectra for all our stars, using the ISAAC spectrometer. We used the band passes of Blum et al. (2003) as well as their continuum points. Effective temperatures were determined by using the following relation  $T_\mathrm{eff} = 4828.0 - 77.5 \times \mathrm{EW(CO)}$ (see Blum et al. 2003), with uncertainties of the order of 100\,K. We apply this method for all our Bulge stars. For the three warmer thick-disk stars, we used the temperature determination from \citet{monaco:11}, which is also where we chose our thick disk stars. They determined \teff\ by imposing an excitation balance for Fe\,{\sc i}  lines.


Surface gravities  were determined in the same manner as in \citet{ryde_schultheis:15}, by using $H$ and $K_s$ band photometry and adopting a mean distance of 8.4\,kpc to the Bulge \citep{GCdistance}.
 In this calculation,  we used $H$ and $K_s$ band photometry from 2MASS \citep[][for the $b=-1^\circ$ and $b=-2^\circ$ fields]{2mass}, Nishiyama (private comm.) and the VVV survey \citep[][for the GC field]{minniti2010}, extinction values from \citet{schultheis2009}, and the bolometric corrections from \citet{houdashelt2000}. We used $T_{\odot}$ = 5770 K, $\log g_{\odot}$ = 4.44, $M_{\rm bol \odot}$ = 4.75, and assumed that our giants are of 1.0 M$_\odot$.  \citet{monaco:11} provide \logg\ for the disk stars, except for the thin disk star BD-012971, whose surface gravity was taken from \citet{origlia_M}.

The metallicities and microturbulence parameters are determined from our high-resolution, CRIRES spectra, see next paragraph \ref{stellarabund}. Our derived stellar parameters are given in Table \ref{starsall_tab}. The fundamental parameters of our comparision star $\alpha$ Boo are taken from \citet{aboo:param}.

\subsection{Stellar abundances\label{stellarabund}}





 

With the stellar parameters (or the fundamental parameters of the star) determined, we can now obtain the stellar abundances of the elements we are interested in from the high-resolution CRIRES spectra. We analyse these using  the software  {\it Spectroscopy Made Easy, SME} \citep{sme,sme_code}, which is a spectral synthesis program that uses a grid of model atmospheres in which it interpolates for a given set of fundamental parameters of the analysed star. In our case we used a grid of MARCS spherical-symmetric, LTE model-atmospheres \citep{marcs:08}. The spectra that are analysed with SME are pre-normalized and the spectral lines which are of interest are marked with masks that determine what parts of the lines (i.e. the entire line), that will be  modelled with synthetic spectra. Continuum masks are used to define regions that SME should treat as continuum regions for an extra linear rectification in predefined, narrow windows around the lines analysed.
SME then iteratively synthesise spectra for a given set of free variables, which could for example be the searched abundance, under a scheme to minimise the $\chi^2$ when comparing with the observed spectra.  The best fit will provide the abundance of the element from the observed spectral lines.  For details, see \citet{sme}. 

In order to obtain accurate abundances, the atomic data of the spectral lines used have to be known. An atomic line-list based on an extraction from the VALD database \citep{vald} was therefore constructed. Due to inaccurate atomic data, and the lack of laboratory measurements, we had to fit the atomic lines to the solar intensity spectrum of \citet{solar_IR_atlas}, by our determining astrophysical $\log gf$ values for, most importantly,  Fe and Si lines. To test these lines we then derive the metallicity and Si abundance for $\alpha$ Boo, which we find to be  within 0.05 dex of the values determined by \citet{aboo:param}. The stellar parameters we use for $\alpha$ Boo  are those from  the detailed investigation of these authors: \teff$=4286\,$K, $\log g =  1.66$ (cgs), [Fe/H]$=-0.52$, $\chi_\mathrm{micro}=1.7\,$\kms, and $\chi_\mathrm{macro}=6.3\,$\kms. Furthermore, we used $^{12}\mathrm C/^{13}\mathrm C = 9$ as determined by \citet{abia:12} and C, N, and O abundances as derived in \citet{ryde:bulb1} and \citet{abia:12}. These are important in order to synthesise the molecular lines in the K band properly.

Since 
the Mg lines are impossible to fit in the solar spectrum, we have determined their line strengths such that they fit the standard giant star Arcturus' flux spectrum \citep{arcturusatlas_II} instead, with a Mg 
abundance from \citet{aboo:param}; [Mg/Fe]$=0.37$. 

In the  syntheses of our stars we have also included a line list of the only molecules affecting this wavelength range, namely $^{12}$CN and $^{13}$CN (J\o rgensen \& Larsson, 1990)\nocite{jorg_CN}.  

In the abundance determination, the microturbulence,  $\chi_\mathrm{micro}$, is important for the spectral syntheses, affecting saturated lines. We have chosen to use a typical value of $\chi_\mathrm{micro}=2.0$ km\,s$^{-1}$ found in detailed investigations of red giant spectra in the near-IR by \citet{tsuji:08}, see also the discussion in \citet{cunha2007}. 
Finally, in order to match our synthetic spectra with the observed ones, we also convolve the synthetic spectra with a `macroturbulent' broadening, $\chi_\mathrm{macro}$, which takes into account the macroturbulence of the stellar atmosphere and instrumental broadening. $\chi_\mathrm{macro}$ is found by using SME with this parameter set free, for a set of well determined lines, for every star.


In  the same manner as described in \citet{ryde_schultheis:15}, the iron (giving the metallicity), Mg and Si abundances were then determined for all our stars by $\chi^2$ minimisation between the observed and  synthetic spectra for, depending on star, $6-9$ Fe, $2$ Mg, and $2-6$ Si lines. 
The iron lines have excitation energies between $3-6$\,eV and line strengths of $\log gf = -4$ to $0.3$.
In contrast to \citet{ryde_schultheis:15}, we do not use the Ca abundances due to unknown systematic uncertainties arising from using the Ca lines in our wavelength region. 
To fit the CN lines that might affect the lines we are interested in, we used typically $10$ unblended CN lines and fit these by letting the either the C or N abundance free (see Table \ref{cn}). We then  assume a given typical value of the N enrichment or C depletion, respectively, in a 'heavily CN-cycled' red-giant atmosphere, which has experienced the first dredge-up along the subgiant-giant branch, converting C into N: [N/Fe]$=+0.53$ or [C/Fe]$=-0.38$ \citep[for details see][]{marcs:08}. The final metallicities and $\alpha$-element abundances we derive, are presented in Table \ref{abund} and the synthesised spectra are shown with the observed ones for a few typical stars in Figure \ref{spectra}.

The uncertainties in the determination of the abundance ratios, for typical uncertainties in the stellar parameters, are all modest, less than 0.15 dex, \citep[see Table \ref{errors} and the discussion in][]{ryde_schultheis:15}. Systematic uncertainties include, for example, non-LTE effects and the continuum placement, but are difficult to estimate. The latter we estimate to of the order of 0.1 dex. We thus estimate a total uncertainty of $\pm0.15$ dex.  To show the  sensitivity of the CN lines to uncertainties in the stellar parameters,  we show the uncertainty in the nitrogen abundance for a given C abundance in Table \ref{errors}, which thus is the abundance change needed to fit the CN lines.

 \begin{deluxetable}{lcccc}
\tablecaption{Uncertainties in the derived abundances due to typical uncertainties in the stellar parameters for a typical star$^a$\label{errors}}
\tablewidth{0pt}
\tablehead{
\colhead{Uncertainty} & \colhead{$\delta \mathrm{[Fe/H]}$} & \colhead{$\delta \mathrm{[Mg/Fe]}$} & \colhead{$\delta \mathrm{[Si/Fe]}$} & \colhead{$\delta \mathrm{[N/Fe]}$} }
\startdata
$\delta T_\mathrm{eff} = +150\,\mathrm{K}$ & $+$0.02 & $-$0.02 & $-$0.10 & $-$0.19\\
$\delta \log g = +0.5$ & $+$0.02 & $-$0.02 & $\pm0.00$ & $+0.02$ \\
$\delta \mathrm{[Fe/H]} = +0.1$ &   & $-$0.08 & $-$0.09 & $-$0.07 \\
$\delta \xi_\mathrm{micro} = +0.5$ & $-$0.10 &  $-$0.07 & $-$0.03 & $-$0.05\\
\enddata
\tablecomments{The stellar parameters used are $T_\mathrm{eff}=3700$~K, $\log g = 1.5$, $\xi_\mathrm{micro}=2.0$~km\,s$^{-1}$, and solar metallicity. The uncertainty in the nitrogen abundance is for a given C abundance, and is the uncertainty we find when fitting the CN lines, see text.}
\end{deluxetable}

  



 
 




\begin{deluxetable}{lccc}
\tabletypesize{\scriptsize}
\tablecaption{Derived abundances$^{a}$\label{abund}}
\tablewidth{0pt}
\tablehead{
\colhead{Star} & \colhead{[Fe/H]} & \colhead{[Mg/Fe]} & \colhead{[Si/Fe]} }
\startdata
GC1  &  0.15 	 & 0.04 & 0.12 \\ 
GC20 &  0.14 	 & 0.10 & 0.04 \\ 
GC22 &  0.04 	 & 0.01 & 0.05 \\ 
GC25 &  $-$0.20  & 0.20 & 0.20 \\ 
GC27 &  0.23 	 & 0.01 & 0.15 \\ 
GC28 & $-$0.04  & 0.07 &$-$0.01 \\ 
GC29 &  0.12 	 & 0.05 &$-$0.06 \\ 
GC37 & $-$0.08  & 0.06 & 0.18  \\ 
GC44 &  0.18 	 & 0.03 &$-$0.18 \\ 
\tableline
bm1-06 &   0.29       &   0.20   &   0.10   \\ 
bm1-07 &   0.08   &   0.10   &  $-$0.03 \\ 
bm1-08 &   0.18    &   0.01   &   0.01   \\ 
bm1-10 & $-$0.22     &   0.25   &   0.10   \\ 
bm1-11 &  0.12     &  $-$0.04 &   0.05   \\ 
bm1-13 &  $-$0.95   &   0.48   &   0.43   \\ 
bm1-17 &  $-$0.83     &   0.59   &   0.54   \\ 
bm1-18 &   0.22    &  $-$0.01 &  $-$0.03 \\ 
bm1-19 &   0.18       &   0.16   &   0.06   \\ 
\tableline
bm2-01  &  0.15  	 &  0.23   &  0.11  \\ 
bm2-02  & $-$0.48  	  &  0.40   &  0.25  \\ 
bm2-03  &  0.26  	  &  0.07   & $-$0.01  \\ 
bm2-05 &  0.01 	  &  0.10   & $-$0.09  \\ 
bm2-06 &  $-$1.17 	  &  0.36   &  0.32  \\ 
bm2-11 &  $-$0.91 	  &  0.42   &  0.36  \\ 
bm2-12 & $-$0.11  	 &  0.07   & $-$0.06  \\ 
bm2-13 & $-$0.16  	  &  0.26   &  0.06  \\ 
bm2-15 &  0.22  	  &  0.04   &  0.04  \\ 
bm2-16 &  0.10  	 &  0.15   &  0.04  \\ 
\tableline
$\alpha$ Boo 		& $-$0.53  & 0.37$^b$ & 0.33  \\ 
BD-012971   		&  $-$0.81  &  0.33  &   0.29  \\ 
142173$^{c}$   		&  $-$0.77  &  0.27  &   0.26  \\ 
171877$^{c}$   		&  $-$0.92   &  0.45  &   0.45  \\ 
225245$^{c}$   		&  $-$1.10  &  0.44  &   0.33  \\ 
313132$^{c}$   		&  $-$0.28   &  0.29  &  0.27   \\ 
343555$^{c}$  		&  $-$0.72  &  0.44  &   0.43  \\ 
HD 787        		&  $-$0.22  &  0.18  &   0.17  \\ 
\enddata
\tablenotetext{a}{$\mathrm{[X/Fe]} = \{\mathrm { \log\varepsilon(X)} - \mathrm { \log\varepsilon(Fe)\}_{star}} - \{\mathrm { \log\varepsilon(X) } - \mathrm { \log\varepsilon(Fe)\}_\odot }$, where $\varepsilon(X)$ is the number density of the species $X$.}
\tablenotetext{b}{Mg 
abundance from \cite{aboo:param}.}
\tablenotetext{c}{Identification number from the Southern Proper-Motion Program \citep[SPM III][]{girard:04}, as given  in \citet{monaco:11}. }
\end{deluxetable}

\section{Results}

In Table \ref{abund} we present our abundances: metallicities, [Fe/H], and the $\alpha$-element abundances of [Mg/Fe] and [Si/Fe]. 
We have normalised our derived abundances to the solar abundances of \citet{solar:sme}: 
$\mathrm { \log\varepsilon(Mg)=7.53}$, $\mathrm { \log\varepsilon(Si)=7.51}$, 
   and $\mathrm { \log\varepsilon(Fe)=7.45}$.
We have redetermined the abundances for the Galactic center stars of \citet{ryde_schultheis:15}, in order to be homogenous with the other fields. They only change slightly, in most cases by much less than 0.1 dex, but in a few cases by more than that, which demonstrates that systematic uncertainties can be of that order for stars of low signal-to-noise, or heavy line density in the spectrum.

\begin{deluxetable}{lccc}
\tabletypesize{\scriptsize}
\tablecaption{Derived abundances for given values of nitrogen and carbon, respectively\label{cn}}
\tablewidth{0pt}
\tablehead{
\colhead{Star} & \colhead{[C/Fe]} & \colhead{or} & \colhead{[N/Fe]} \\
\colhead{ } & \colhead{[N/Fe] = +0.53} & \colhead{} &
\colhead{[C/Fe] = -0.38 } 
}
\startdata
GC1  &  0.19 	&	& 	1.72	\\
GC20 &   0.00 	&	& 	0.96	\\	
GC22 &   0.12 	&	&	1.49	\\
GC25 &   0.34 	&	& 	1.71	\\	  
GC27 &   0.16 	&	&	1.58	\\
GC28 &  0.06 	&	&	1.29	\\
GC29 &   0.14 	&	& 	1.55 	\\
GC37 &  0.15 	&	&  	1.41 	\\
GC44 &   0.08 	&	& 	1.57	\\
\tableline
bm1-06 &    0.04    & &	1.53	\\
bm1-07 &   $-$0.45 &  &	1.06	\\
bm1-08 &     $-$0.06 & &	0.96	\\
bm1-10 &   0.01    & &	1.18	\\
bm1-11 &    $-$0.01 & &	1.20	\\
bm1-13 &    0.00    & & 0.71		\\
bm1-17 &    0.21    &  &	1.27	\\
bm1-18 &     $-$0.00 & & 1.29		\\
bm1-19 &    0.09    &  &	0.90	\\
\tableline
bm2-01  &   0.06  	&	&	1.03    \\
bm2-02  &   $-$0.03 	& &  1.02	    \\
bm2-03  &     0.17  &		&	1.48      \\   
bm2-05 &     0.11  	&	&	1.67  \\
bm2-06 &    $-$0.30  &	&  0.49   \\ 
bm2-11 &  	  $-$0.10 & 	&  0.85  	\\
bm2-12 &    0.06  	&	&   1.22 	\\ 
bm2-13 &    0.05  	&	& 	1.17	\\ 
bm2-15 &    0.04  	&	&  1.31  	\\      
bm2-16 &     0.25  	&    &  1.58	\\
\tableline
BD-012971 &    $-$0.03 & &	0.95	 \\
142173  &      0.02  	&  &	0.99	\\ 
171877  &      0.05  	&  &	0.97	\\
225245  &     0.08    & &	1.02 	\\   
313132  &      0.38   &  &	1.52 \\
343555  &     $-$0.01  &	  &	0.96 	\\   
HD 787  &    $-$0.08  &	  &	0.82 	\\ 
\enddata
\tablecomments{We have normalised our derived abundances to the solar abundances of \citet{solar:sme}: 
$\mathrm { \log\varepsilon(C)=8.39}$, $\mathrm { \log\varepsilon(N)=7.78}$, and and $\mathrm { \log\varepsilon(Fe)=7.45}$.
}
\end{deluxetable}

We have not determined the C and  N-abundances for our stars since we do not have any CO lines in our wavelength range. However, we have numerous CN lines that we need to fit in the spectral synthesis in order to take blends into account properly. We have therefore determined the C and N abundances such that these are synthesised properly. Since the C and N abundances are degenerate, we have determined one of them for a given typical abundance of the other. This  means that our carbon or nitrogen abundances are not necessarily the stellar abundances, but rather merely the values we need in combination with the given N or C abundances, respectively, to fit the CN lines. 

In the second column in Table \ref{cn} we thus provide the [C/Fe] abundance that is required to fit the CN lines for a given  typical N enhancement for heavily CN-cycled red giants, [N/Fe]$=+0.53$ \citep{marcs:08}. We see that the carbon abundances, in this case, are more than scaled solar, which is not expected for these type of giants \citep{smith:13}. In the third column in the Table, conversely, the  [N/Fe] abundance that is required to fit the CN lines for a given  typical C depletion, [C/Fe]$=-0.38$ \citep{marcs:08}, are provided. In order to fit the CN lines, in this case, unexpectedly high intrinsic nitrogen abundances are needed. There thus seems to be an indication that we need a high [C+N/Fe] abundance in these bulge stars, but also our thick disk stars. 
We note that \citet{rich:07,rich:12} measure the C, N, and O abundances in M giants in the inner Bulge and find typical red giant values. In \citet{melendez:2008,ryde:10}  the C, N, and O abundances are also measured from CO, CN, and OH lines, which we cannot do in the present spectra, and they find only slightly enhanced [C+N/Fe] for K giants. We find a more extreme change of the C and N abundances, more resembling Arp4203 in \citet{ryde:10}, which is a star more resembling those in this paper. We find no correlation with $K_0$ magnitude, but there is a clear correlation with \teff, in the sense that the lower the temperature, the higher the C or N enhancement. In Table \ref{errors} we show that this large discrepancy can not be explained by an uncertainty in, e.g., \teff. 
We do not discuss this further, since we have not measured these abundances, but merely infer them from forcing the CN lines to fit the observations. Determining the C, N, and O abundances of these stars or more inner bulge stars would be rewarding.



We plot the [Mg/Fe] and [Si/Fe] trends  as a function of the metallicity, [Fe/H], in Figure \ref{afe1}  separately for each field: the Galactic center, the $(l,b)=(0,-1^\circ)$, and for the $(l,b)=(0,-2^\circ)$ fields.  In Figure \ref{afe2}, we show the trends for all our observed stars in the entire region (within a projected distance of 300 pc) together. In this Figure, we also include the mean abundance ratios of the two populations of Terzan 5 at $(l,b)=(3.8^\circ$,$+1.7^\circ)$ from \citet{origlia:11}.  From the Figures we see that the $\alpha$-abundance trends, here represented by the [Mg/Fe] and the [Si/Fe] versus [Fe/H] trends, are very similar in the three inner Bulge fields.  Plotting the trend for all our stars in the inner 300 pc region strengthens this finding.

In the Figures showing the inner Bulge trends, we have also included the trends outlined by the abundances derived from micro-lensed dwarfs from the `outer' Bulge $(|b|>-2^\circ$) as derived by \citet{bensby:13}.  The [Mg/Fe] and [Si/Fe] trends in all three inner regions, from the Galactic center to a latitude of $b=-2^\circ$, follow the outer bulge trends, within uncertainties, with low values of [$\alpha$/Fe] at [Fe/H]$>-0.2$. From these plots it is evident that we can not claim that our [$\alpha$/Fe] trend in the inner 300 pc is particular compared to the `outer' Bulge, $b>-2^\circ$. Our thick disk trend also follow the micro-lensed dwarf trend, which gives confidence in the analysis method and shows that we are on the same scale, with room for only small systematic uncertainties.

The metallicity of the $b=-2^\circ$, $b=-1^\circ$, and Galactic center fields ranges from $-1.2<\mathrm{[Fe/H]}<+0.3$, $-1.0<\mathrm{[Fe/H]}<+0.3$, and $-0.2<\mathrm{[Fe/H]}<+0.3$, respectively. The two Southern fields start picking up some metal-poorer stars, some at very low metallicities (even below [Fe/H]$=-1$). The spread in metallicities we find are larger than what has been found earlier: \citet{rich:07,rich:12} find a dispersion of approximately 0.1 dex around [Fe/H]=$-0.05$ to $-0.15$ and  \citet{cunha2007} find a total spread of $0.16$ dex around [Fe/H]=$+0.14$ for the luminous giants and supergiants in the Central Cluster located within 2.2 pc of the Galactic center.

The metallicity distributions for the different fields we find in this investigation, are presented in Figure \ref{mdf}, with the upper three panels showing the three inner fields, and the lowest panel showing the combined metallicity-distribution histogram. The global mean-metallicities of all the observed stars in each of the inner Bulge regions are, from the Galactic center and outwards, $<[\mathrm{Fe/H}]> = +0.06$, $-0.10$, and $-0.21$, respectively.   For comparison, we can compare our mean metallicities with \citet{rojas:15} who find a mean metallicity of approximately $<$[Fe/H$>$]$=-0.17$ for their fields at $(l,b)=(1^{\circ},-4^{\circ})$ and $(0,-6^{\circ}$), i.e. at slightly lower latitudes. Our mean metallicities within the inner $2^\circ$ gives a formal gradient of $-0.06\,\mathrm{dex}/100\,\mathrm{pc}$ or $-0.08$\,dex/degree. This number should be handled with caution due to the very low number of stars. Also, there might be undetermined selection effects, which prevents us from  selecting metal-poor stars, due to the large and variable extinction in the Galactic center field. More stars analysed in the same maner as done here are needed to determine the metallicity distributions in these fields, in order to derive a gradient in the inner bulge.




\section{Discussion}

Here, we will discuss our results concerning the $\alpha$-abundance trends, and the metallicity distributions and gradients in the inner Bulge. 


\subsection{$[\alpha/\mathrm{Fe}]$ trends with metallicity}

\subsubsection{Observational results}

\begin{figure*}
  \centering
	\includegraphics[width=\hsize]{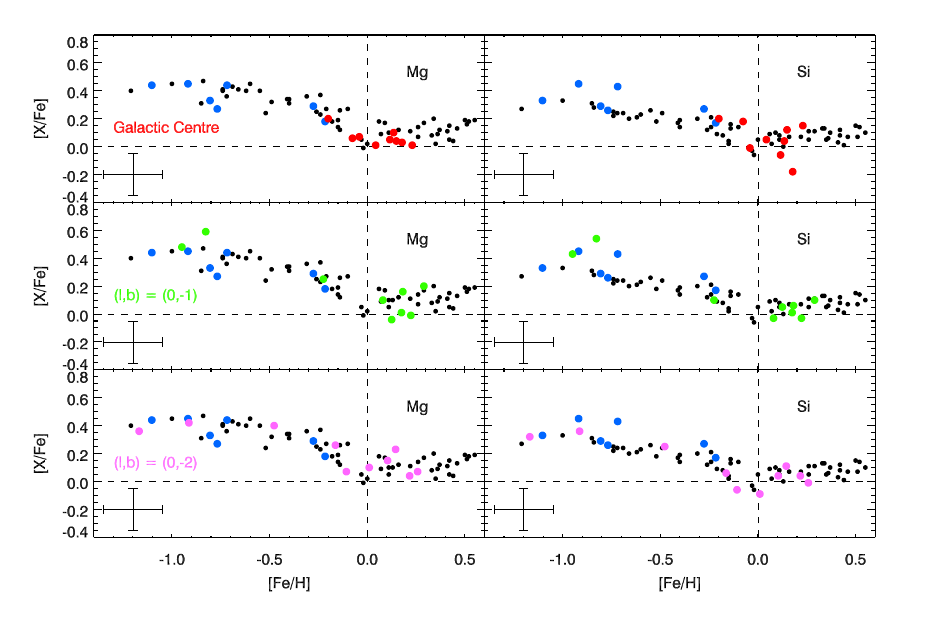}
	\caption{Abundances trends of [Mg/Fe] and [Si/Fe] versus [Fe/H]. We estimate an uncertainty of $\pm 0.15$ dex in [X/Fe] and [Fe/H]. In the upper panel we present the trends for the Galactic center field (red dots). The middle panel shows our derived trends from the $(l,b)=(0,-1^\circ)$ field (green dots). In the lower panel we present the trends for the $(l,b)=(0,-2^\circ)$ field (purple dots). In all the plots, the blue dots mark our local disk star measurements, and the black small dots are the abundances determinations based on micro-lensed dwarfs by \citet{bensby:13}. } 
	\label{afe1}
\end{figure*}

\begin{figure*}
  \centering
	\includegraphics[width=\hsize]{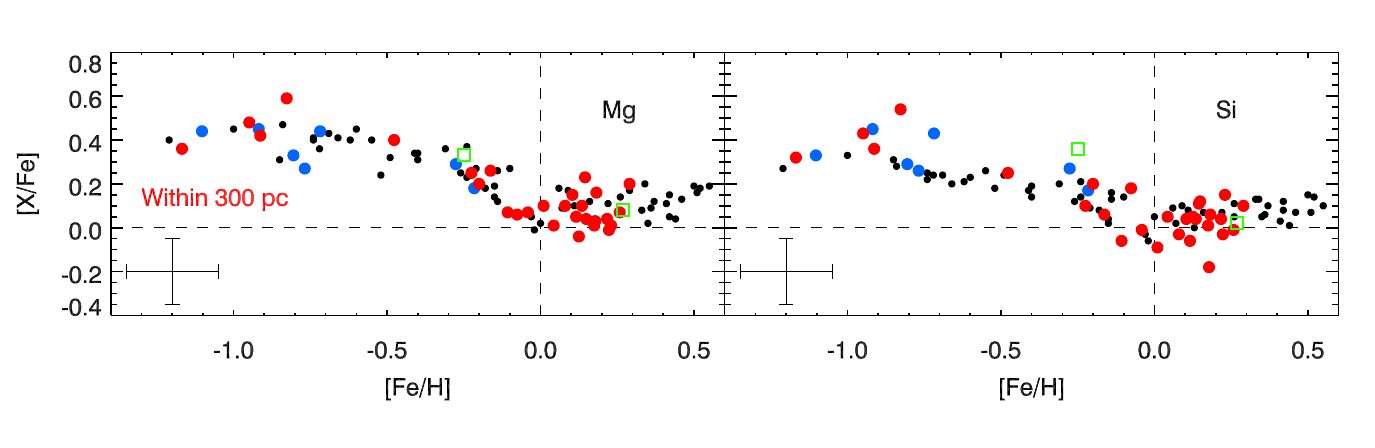}
	\caption{Abundances trends of [Mg/Fe] and [Si/Fe] versus [Fe/H]. We estimate an uncertainty of $\pm 0.15$ dex in [X/Fe] and [Fe/H]. The Figure shows our derived trends for all stars in the three fields together (red dots). The mean abundance ratios of the two populations of Terzan 5 at $(l,b)=(3.8^\circ$,$+1.7^\circ)$ from \citet{origlia:11} are shown with green squares. For the inner globular cluster Liller 1  at $(l,b)=(354.8^\circ$, $-0.2^\circ)$, \citet{origlia_GC4} find [Fe/H]$\sim-0.25$ and [$\alpha$/Fe]$\sim+0.3$], which fits nicely with the general trend of the Rich et al. (2007, 2012) field stars.  In all the plots, the blue dots mark our local disk star measurements, and the black small dots are the abundances determinations based on micro-lensed dwarfs by \citet{bensby:13}.
} 
	\label{afe2}
\end{figure*}


Looking at the Mg and Si trends, we observe very similar [$\alpha$/Fe] versus [Fe/H] trends for all three inner fields, given the uncertainties. They are also very similar to the trends found by, for example,  \citet{bensby:13}, who analysed micro-lensed dwarfs further out in the Bulge ($b>-2^\circ$). This is the sample we have compared with in the Figures. This suggests a quite homogeneous Bulge when it comes to the star-formation history. We can conclude that there are no detectable gradients, to the level of the uncertainties of the present observations, in the [$\alpha$/Fe] levels with latitude. 

It is interesting to note that \citet{johnson:11} find an identical [$\alpha$/Fe] enhancement trend in Plaut's window at $(l,b)=(-1^\circ,-8.5^\circ$) as in Baade's window at $(l,b)=(1^\circ,-4^\circ)$, indicating a lack of an [$\alpha$/Fe] gradient in the outer Bulge ($-9<^\circ b<-4^\circ$. \citet{gonzalez:11} also find no difference in the  [$\alpha$/Fe] in fields at $-12^\circ <b<-4^\circ$. \citet{rich:07,rich:12} also find no gradient from $b=-4^\circ$ to $-1^\circ$. J\"onsson et al. (in prep.) find a similar conclusion for fields between $-6^\circ<b<-3^\circ$. These latter trends are also similar to those in our inner fields $-2^\circ<b<0^\circ$ presented here. {\it We, therefore, conclude that the lack of an [$\alpha$/Fe] gradient further out extends all the way into the Galactic centre}.
 
In Figure \ref{afe1} we also plot the mean [Mg/Fe] and [Si/Fe] abundances of the two populations in the complex globular cluster Terzan 5 located in the inner Bulge, $1.7$ degrees North of the Galactic Plane, from \citet{origlia:11}. The [Mg/Fe] abundances follow our trend identically, whereas the [Si/Fe] abundances are close, with the low-metallicity component being marginally higher than our trend. However, the metal-rich population with a mean at [Fe/H]$= +0.3$ lies at the high-metallicity end of our distribution. We also note that, in contrast to Terzan 5, we clearly miss a distinct metal-poor population in our Galactic centre field. \citet{origlia:11} suggest that there might be a common origin and evolution of field populations in the Bulge and Terzan 5, the latter perhaps being a relic building block of the Bulge.  The similarity in the [$\alpha$/Fe] trends are not inconsistent with this idea.

From Figure \ref{afe2} we see that the spread in metallicities in the  two outer fields span the range similar to what has been found in the `outer' Bulge, \citep[see e.g.][]{bensby:13,rojas:15}, except that we do not find as high metallicity stars as they do. We do not know the cause for the difference in the high-metallicity end. 
One possibility is the large difficulty in measuring metallicities, due to the increased number of blends, known and unknown.  Not taking these into account will systematically lead to too high metallicities. The near-IR wavelength region is not affected as much due to a lower line density. We  note that \citet{uttenthaler:15} also find few M-type giants with super-solar metallicities.

Our Galactic centre field, however, shows only  the metal-rich component, also found by \citet{ramirez:00} and \citet{cunha2007}, even though they found a narrower distribution. We thus find that the metal-poor component seems to disappear at the Galactic centre. This could either reflect a true absence of metal-poor stars or it could be due to a selection bias. The interstellar extinction in the Galactic centre is extreme and variable and its estimates are not accurate enough, which means that it is   very difficult to define an metallicity-unbiased sample of stars. Due to this and due to the low number of stars we, therefore, do not want to draw too much conclusions from this fact.






\citet{babu:10,gonzalez:11} find that the metal-rich component is more concentrated to the plane and progressively disappears with increasing latitude, whereas the metal-poor component shows the same trends and distributions, and kinematics at all the regions investigated, namely $-12^\circ <b<-4^\circ$. We thus find that this is not the case in the Galactic centre. 

Before, discussing the metallicity gradients in the inner Bulge further, we will discuss our Galactic chemical evolution modelling of the inner Bulge.

\subsubsection{Galactic Chemical Evolution modelling of the inner Bulge}

\begin{figure*}
  \centering
	\includegraphics[width=\hsize]{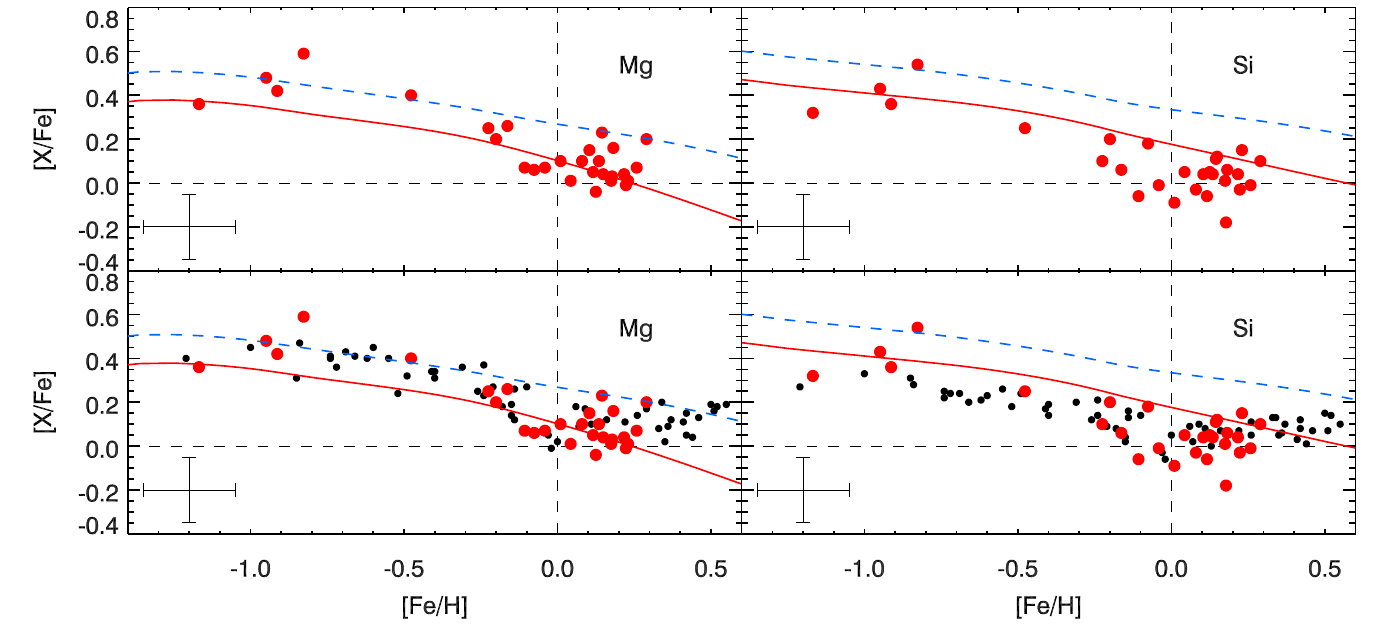}
	\caption{Predicted and  observed [Mg/Fe] and [Si/Fe] vs. [Fe/H] for the inner 500 pc of the Galactic bulge. Our derived abundance ratios for all stars in the three fields together are shown with red dots. The predictions are represented by the two curves: the dotted curve represents the predictions of the model of \citet{grieco:15}  (model labelled (ii) in that paper), whereas the continuous line is the prediction of the same model except for the yields of Mg and Si which have been decreased by a factor of 1.35 relative to \citet{grieco:15} (see text). In the lower panel we have added  the abundances determinations based on micro-lensed dwarfs in the `outer' bulge by \citet{bensby:13}, marked with black small dots.} 
	\label{vale}
\end{figure*}

%
%

 We have modelled the Galactic chemical evolution of the inner Bulge by using the classical-bulge model adopted in \citet{grieco:12} for the so-called metal-poor population, recently used to model the Galactic Center by \citet[][their model {\it (ii)}; see their Figure 8]{grieco:15}.
In this model we assume a very efficient star formation (with efficiency larger than a factor of 20 relative to the solar vicinity), an  IMF with more massive stars than normally adopted for the solar neighbourhood \citep{chabrier:03}.
We have assumed that the bulk of  bulge  stars formed during  a fast episode of gas infall occurring on a timescale of  $\sim 0.3$ Gyr.  The adopted stellar yields are the same as in \citet{romano:10} for massive, low- and intermediate mass stars and supernovae Type Ia. These yields have been tested on solar neighbourhood stars and can well reproduce the observed [Mg/Fe] and [Si/Fe ] trends in those stars.

  In Figure \ref{vale} we show the predictions of this model for the inner 500 pc relative to the abundance ratios [Mg/Fe] and [Si/Fe]. We report two curves where the only difference are the yields of Mg and Si from massive stars. One curve represents the best model of \citet{grieco:15} where the yields of Mg and Si from massive stars are taken from \citet{kobayashi:06}, whereas the other curve represents a model where the Kobayashi et al. yields of Mg and Si have been decreased by a factor of 1.35. This is allowed by the fact that theoretical yields have uncertainties. Therefore, the two curves mark the theoretical uncertainties in the stellar yields. The agreement with the [$\alpha$/Fe] vs. [Fe/H] trend is good for both elements, confirming our previous conclusions in \citet{grieco:15} about the fast formation of the bulk of bulge stars.

It is worth noting that the fraction of SNe Ia in the IMF ($\sim 0.18$), adopted in the model, well reproduces the SNe Ia rate quoted by \citet{li:11}. In particular, our predicted rate is $0.2-0.3$ SNe Ia per century.
This value is larger than previous estimates \citep[e.g.][]{schanne:07}, who quoted 0.03 SNe Ia per century but pointed out that their value for the Type Ia SN rate is an order of magnitude lower than expected to explain the large positron injection rate into the Galactic Centre region, as observed by INTEGRAL \citep[][and references therein]{schanne:06}. The more recent  value of \citet{li:11}, instead, is between 0.25 and 0.3 SNeIa per century, in very good agreement with our predicted one.

The inner 300 pc includes the Central Molecular Zone (CMZ), which is a rich environment with a recent intense star formation \citep[within 100,000 yrs;][]{yusef:09}, massive stars, and is home for three of the most massive young clusters in Milky Way. However, most stars in this region are nevertheless old, with ages larger than 9 Gyr \citep{genzel:06}. To reproduce the star-formation rate (SFR) at the present time,  \citet{grieco:15} over-impose a recent star burst fitting literature SFR. The origin of the gas can either be from merger processes or from accretion of inner disk gas (galactic bar). This extra burst does however not affect the [$\alpha$/Fe] trends.





\subsection{Metallicity gradients in  the inner Bulge}

\begin{figure}
	\includegraphics[angle=90,width=\hsize]{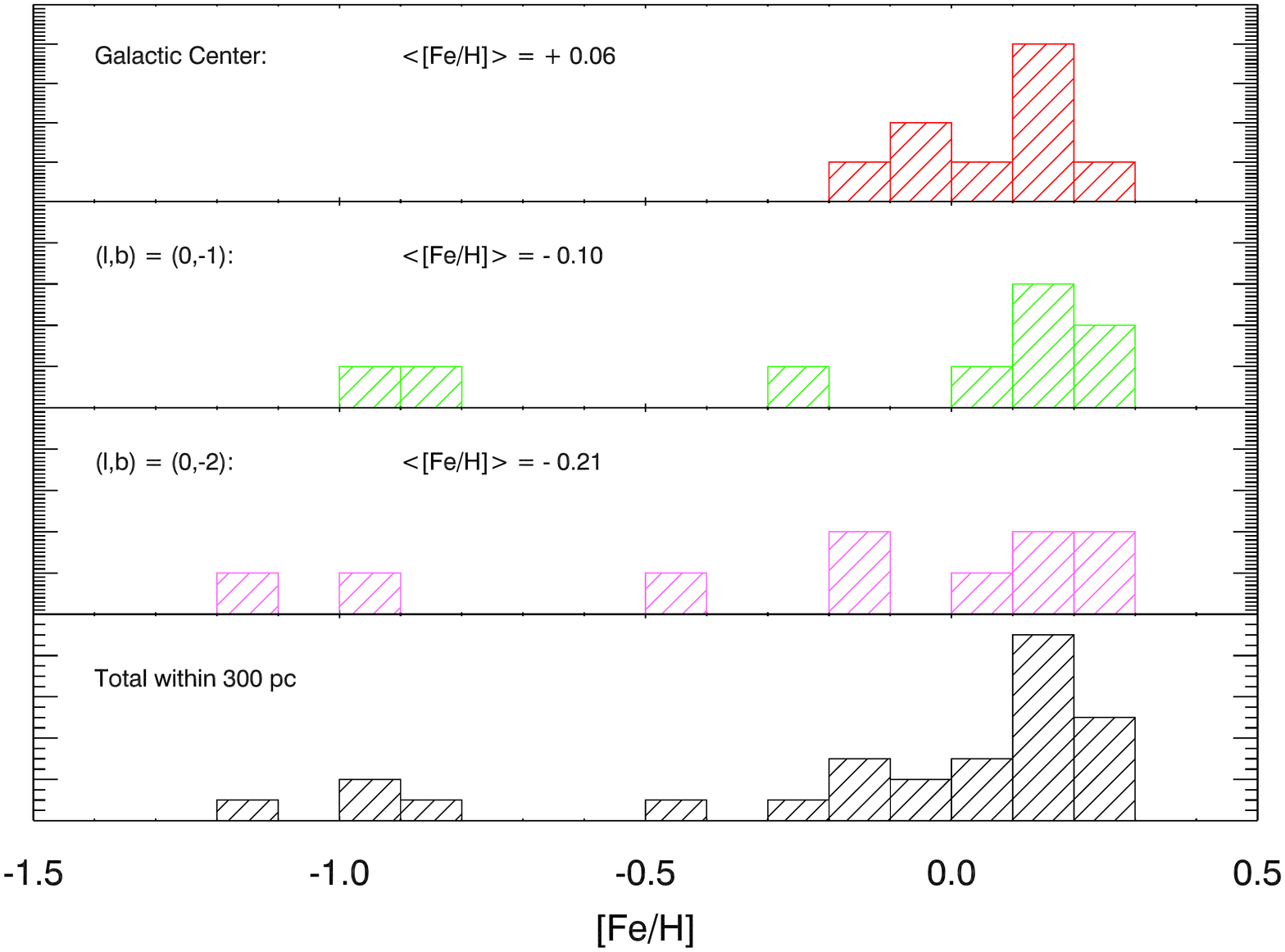}
	\caption{Metallicities of our stars in the Galactic centre (top panel) and the stars in $b=-1^\circ$ and $b=-2^\circ$ fields, below. The lowest panel shows all the stars  within $b=-2^\circ$ of the Galactic center in one plot. The global mean of the stars for the three fields are indicated.} 
	\label{mdf}
\end{figure}

The three top panels of Figure \ref{mdf} show the metallicity distributions for our three inner Bulge fields. The fact that we have very few stars in every region, means that we hardly can talk about metallicity distribution functions. Also, there might be a selecition effect against metal-poor stars in the Galacitc center field, that we have not quantified. 

With these caveats in mind, our data shows that the distributions progressively get more metal-rich in the mean, as we approach the Galactic centre. The data are consistent with a vertical metallicity gradient. Taking the straight mean of the metallicities of the giants in every field, we arrive at a metallicity gradient of $0.08$ dex/degree or $0.06$ dex/100pc\footnote{For a distance of 8 kpc to the Bulge, $1^\circ$ corresponds to 140 pc}. The metal-rich component seems to be stable whereas the metal-poor stars are absent in the Galactic centre field. One should, however, be cautious due to the few stars yet available; given the number of stars in the Galactic centre field (9) and the number of stars with [Fe/H]$<-0.5$ in the other fields (4/19), it is not
so unlikely that metal-poor stars were missed in the Galactic center field only because
of the small number of observed stars. Taking the statistics from the two
other fields as a basis, we get $(1-(4/19))^9=0.12$, which is not a
negligibly small probability that no star with [Fe/H]$<-0.5$ is found if
nine stars are being observed. 

Whether there are two components with different proportions at different latitudes or not, is still debated.  It should be noted that if there were a clearly distinct metal-poor population with higher scale height and significant numbers, that population should have presented itself clearly in the BRAVA survey \citep{brava:1,kunder:12}, which is did not.  The cylindrical rotation signature of BRAVA is stable through $b<-8^\circ$, which argues for a single population with an intrinsic gradient.

It is qualitatively well established that there exists a metallicity gradient in the outer bulge along the minor axis from $-12^\circ$ to $-2^\circ$  of the order of $-0.04$ to $-0.10$ dex/deg \citep[e.g.][]{minniti:95,zoccali:2003,zoccali:08,johnson:11,johnson:13,gonzalez:13,rojas:15}. Also, recent results from the GIBS survey show a clear gradient \citep{gibs}.  More specifically, \citet{gonzalez:13} find, for example, a gradient of $-0.04$ dex/degree 
from a global photometric metallicity map of the entire $|b|>2^\circ$ Bulge. This is close to what  \citet{grieco:12} predicted based on a chemodynamic model after a initial collapse, 
 of the inner 500 pc region of the Bulge.  
 \citet{rojas:15} study 1200 Bulge stars from the {\it Gaia}-ESO survey in 5 fields, the closest being at a latitude of $-4^\circ$ and find a gradient of $-0.05$\,dex/degree ($-10^\circ<b<-6^\circ$), which however vanishes at $-6^\circ<b<-4^\circ$. When extrapolating their variation of the two main components from $-4^\circ$ inward, they predict a dominance of the metal-rich component and therefore a weakening of the gradient, which mainly reflects the relative strengths of the two components. This is also seen in the investigation by \citet{massari:14} of field stars around Terzan 5 ($b=+1.7^\circ$). Furthermore,  \citet{johnson:11,johnson:13} determine a gradient of between $-0.06$ and $-0.10$\,dex/degree for the region $-12^\circ<b<-4^\circ$ and \citet{zoccali:08} determined a gradient of the order of $-0.08$\,dex/degree for the region $-6^\circ<b<-4^\circ$. This is also what we find, for our inner fields ($-2^\circ<b<-0^\circ$). Based on our data and these numbers there could be a metallicity gradient of the same order of magnitude extending all the way into the centre. However, in order to state something firmly, we would need many more stars analysed in the same maner and we would need to quantify the selection effect against metal-poor stars, if any. Thus, a gradient may be present all the way to the center, but this finding requires further confirmation.

Some earlier investigations  of the metallicity gradient within 4 degrees found evidence for a flattening-out of the gradient, with a narrow peak at [Fe/H]$=-0.2$ to $-0.1$. \citet{ramirez:00} and \citet{babusiaux:14} find no evidence of a gradient along the minor axis. Furthermore, \citet{rich:07} studied detailed abundances of M-giants in a field at $(l,b) = (0^{\circ},-1^{\circ})$, while \citet{rich:12} studied  fields located at lower latitudes: $(l,b) = (0^{\circ},-1.75^{\circ})$ and $(l,b) = (1^{\circ},-2.65^{\circ})$. They find a narrow iron-abundance distribution of [Fe/H]=$-0.15\pm0.1$ for all three fields,  indicating a lack of any major vertical abundance-gradient.  However, at the Galactic Centre, \citet{ramirez:00} and \citet{cunha2007} find a narrow metal-rich population at [Fe/H]$\sim+0.1$ based on K band spectra. Recently, \citet{ryde_schultheis:15} also find a metallicity peak in the Galactic centre at [Fe/H]$\sim+0.1$, but with a slightly larger spread, also from K-band spectra. These are the same data as we present here, however, slightly updated. \citealt{carr2000}, \citealt{ramirez:00}, and \citealt{davies2009} analysed high-resolution spectra  of supergiant stars in the Galactic centre finding near-solar metallicity. Going North, \citet{massari:14} find a  double peak
distribution with peaks at slightly sub-solar and super-solar metallicities with a spread of $\pm 0.25$ dex at $b=+1.7^\circ$. Recently, \citet{schultheis:15} find metal-poor stars beyond 50 pc from the Galactic Center, indicating the presence of
a metal-poor population.   More measurements are clearly necessary to establish the nature of the metallicity gradient, if there is one in the inner regions. 

The detailed studies of \citet{rich:07,rich:12} of M giants at $R=25,000$ in the the $-4^\circ < b <-1^\circ$, mentioned above, probe the same region as our and uses the same type of objects, even though our M giants lie at $K_0\sim9\pm0.5$ in the Galactic center and $K_0\sim10\pm1$ in the Southern fields, whereas theirs lie a bit brighter at $K_0\sim8\pm0.5$. They find a narrow iron-abundance distribution of [Fe/H]=$-0.15\pm0.1$ for all three fields,  but we find a broader distribution. This discrepancy has to be investigated. In order to understand the origin of this difference, we are currently making an differential analysis of these two samples, and investigating the selection of the observed giants in each paper, including how interstellar extinction is traced. 


What causes the metallicity gradient in the Bulge is a matter of debate. Early on, a gradient was though to be an evidence for a classical bulge, formed rapidly by an initial collapse, or through accretion of substructures and/or large violent merger events in the $\Lambda$CDM scenario. However, it has been shown that a gradient can occur also in a purely secular formation scenario; an intrinsic metallicity gradient could, for example,  be produced by the bar due to an original disk gradient \citep{gerhard:13}, or may be caused by winds. 
It is interesting to note that other Bulges in spiral galaxies have measured vertical metallicity gradients \citep[see discussion in][]{rich:12,proctor:99,gonzalez:15}.

\section{Conclusions}

How large spiral galaxies and their various structures are formed is not know today \citep[see e.g.][]{dokkum:13}. An important step on the way to the clarifying this issue, is to study the Milky  Way bulge, which  is a major component of our Galaxy, a galaxy which we can study in detail. Important observables in this context are the [$\alpha$/Fe] trends as a function of metallicity and the latitudinal metallicity-gradients. In this paper, we have presented new abundances and gradients of 28 M-giants located in the inner 300 pc of the Milky Way bulge. We observed the giants at high spectral resolution in the K band, in order to avoid the extreme visual extinction.

Our [Mg/Fe] and [Si/Fe] abundances as a function of metallicity show similar behaviour in all three fields, from the Galactic centre out to a latitude of $b=-2^\circ$. The abundance trends also follow the trends of stars at lower latitudes, in the `outer' Bulge: The abundance ratios are constant up to [Fe/H]$\sim -0.3$, after which they turn down reaching close to solar-values at [Fe/H]=0. The ratios then level off at values slightly above scaled-solar values, up to a metallicity of [Fe/H]$=+0.3$. 

We thus find a lack of  an [$\alpha$/Fe]-gradient in the inner Bulge region, similarly to what has been found for the rest of the Bulge. We thus conclude that the lack of an [$\alpha$/Fe] gradient further out extends all the way into the Galactic centre, which means  a rapid formation scenario and a homogeneity of the enrichment process \citep{rich:12}.

Our Galactic chemical evolution modelling of the inner 300 pc of the Bulge shows that our observed abundance ratios and the metallicity distributions, are compatible with this region having  experienced a main early, rapid ($0.1-0.7$ Gyr), strong burst of star formation, with a high star-formation efficiency. Also, it indicates a need for an IMF skewed to relatively more massive stars compared to the solar vicinity.  The inner 300 pc includes the Central Molecular Zone (CMZ), with a  recent (second burst of) intense star formation. This second burst can not have been trigged by  more than a modest episode of gas infall or accretion of gas induced by the bar \citep{matteucci:15}. This episode does not affect the abundance patterns, however.

Our data shows a trend where the fraction of metal-poor stars increases the further away from to the Galactic Centre one looks. 
In the light of the fact that \citet{rich:07,rich:12} find no metal-poor stars in the inner $4^\circ$, and since we only have very few stars yet observed in every region, and since there might be selection effects in the Galacitc center field due to extreme and variable extinction, often not precisely known, we can not conclude anything about the inner metallicity gradient. More stars in every field is thus needed to draw firm conclusions, which we retrain from doing yet.   A metallicity gradient is well established in the Bulge at $|b|>2^\circ$ from several different studies. Whether this gradient continues all the way into the Galactic centre or not, our data can not tell for certain. What we do find is a quite a large range of metallicities in the inner Bulge. 

To summarise, we have presented the first study to connect old stars in the Galactic Center with the Bulge, using high-resolution spectroscopy, stars of similar luminosities, temperatures, and gravities, and using the same analysis methods, including the same spectral region. It firmly argues for the center being in the context of the Bulge rather than very distinct.  One may tentatively conclude that the very center was not built by the infall of halo or even bulge-like globular clusters, based on metallicity.  The $\alpha$-element trends with metallicity are very similar for the different regions.  Only the metallicity distribution seems to change with latitude. This could be expected if the metal-rich component is defined by the boxy/peanut bulge, which increases its fractional importance toward the centre. The boxy/peanut bulge is the inner structure of the bar, originating from  vertical instabilities in the disk. This stellar population is, therefore,  expected to be similar to the inner disk \citep{sanchez:15}. However, as mentioned earlier, a  distinct metal-poor population should have presented itself clearly in the BRAVA survey \citep{brava:1,kunder:12}, which it did not.

\acknowledgments
We would like to thank  Livia Origlia for very fruitful and insightful discussions. N.R. 
acknowledges support from the Swedish Research Council,
VR (project number 621-2014-5640), 
Funds from Kungl. Fysiografiska S\"allskapet i Lund. 
(Stiftelsen Walter Gyllenbergs fond and M\"arta och Erik Holmbergs donation). 
F.M. and V.G. acknowledge financial support
from PRIN MIUR 2010-2011, project \textquotedblleft The Chemical and dynamical
Evolution of the Milky Way and Local Group Galaxies\textquotedblright, prot.
2010LY5N2T. RMR acknowledges support from grant AST-1413755 from the National science foundation, NSF. This work is based on observations collected at the European Southern Observatory, Chile, program number  089.B-0312(A)/VM/CRIRES and 089.B-0312(B)/VM/ISAAC. This publication makes use of data products from the Two Micron All Sky Survey, which is a joint project of the University of Massachusetts and the Infrared Processing and Analysis Center/California Institute of Technology, funded by the National Aeronautics and Space Administration and the National Science Foundation.


{\it Facilities:} \facility{VLT:Antu}, \facility{VLT:Melipal}.




\bibliographystyle{apj}

\end{document}